\documentclass[prb,preprint]{revtex4-1} 

\usepackage{amsmath} 
\usepackage{amsfonts}
\usepackage{graphicx}
\usepackage{color}
\usepackage{float}
\usepackage{leftindex}
\usepackage{gensymb}

\begin{document}

\title{Unobserving the Moon: the spurious possibility of orbital decoupling due to solar neutrino Arago spot}

\date{April 1st 2024}

\author{Henrik Viitasaari} 
\author{Oskari Färdig}
\author{Joona H. Siljander}
\author{A. Petrus Väisänen}
\author{Aapo S. Harju}
\author{Antti V. Nurminen}
\author{Jami J. Kinnunen}
\email{jami.kinnunen@aalto.fi}
\affiliation{Department of Applied Physics, Aalto University School of Science, FI-00076 Aalto, Finland}

\begin{abstract}
The Arago spot is an intensity maximum at the center of a shadow created by constructive interference of diffracted waves around a spherical object. While the study of diffraction patterns usually concerns visible light, De Broglie's wave nature of matter makes diffraction theory applicable for particles, such as neutrinos, as well. During a solar eclipse, some of the neutrinos emitted by the Sun are diffracted by the Moon, resulting in a diffraction pattern that can be observed on Earth. In this paper we consider the theoretically emerging solar neutrino Arago spot as a means to measure the location of the Moon with high accuracy and consider its implication on the orbit of the Moon given Heisenberg's uncertainty principle. Our results indicate that the Moon is not at immediate risk of orbital decoupling due to the observation of a solar neutrino Arago spot.
\end{abstract}

\maketitle 

\section{Introduction}

Archimedes once famously said: "Give me a lever and a place to stand and I will move the Earth". As Archimedes' ask was hopelessly implausible, in this paper we study the possibility of moving the Moon without such apparatus. Specifically, we study the possible effect of the quantum mechanical uncertainty principle on the Moon's momentum when its spatial location is determined very accurately. We propose a novel means of said determination via neutrino astronomy, namely using the phenomenon of Arago, or Poisson, spot, and analyze the probability for the Moon being thrown off its current orbit. We believe it is imperative that this possibility is made known so as to prevent anyone from accidentally observing the Moon too accurately.

\section{Solar eclipse 2024}

\begin{figure}[ht]
    \centering
    \includegraphics[width=0.5\columnwidth]{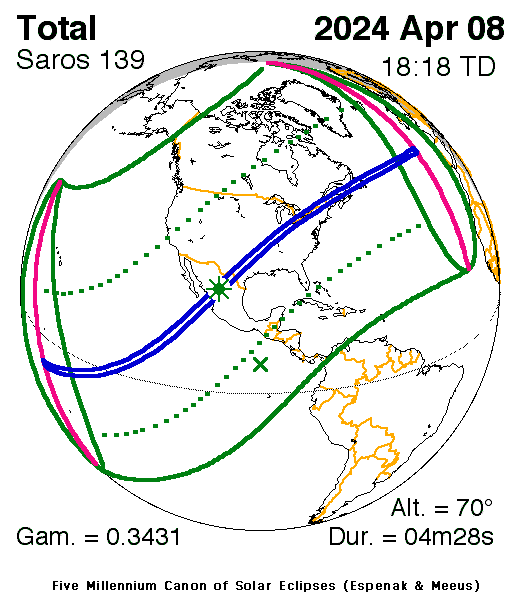}
    \caption{Route of the moon's shadow on 28.4.2024. Total eclipse shown in blue and partial eclipse shown in green. Reproduced with permission.\cite{route}}
    \label{fig:moon-route} 
\end{figure}

Solar eclipses happen when the Moon passes the Sun, casting a shadow on Earth. Such an event will take place on 8.4.2024 \cite{eclipse-times}. The shadow of the moon will cross through North America as seen in Figure \ref{fig:moon-route}. The route of the total eclipse is shown with the dark blue lines. The green lines represent the area where a partial eclipse can be seen. The eclipse starts at 15:42 UTC, when the shadow of the Moon can be seen by observers at the Pacific ocean. The greatest central duration (the time during which the Sun is completely blocked) is 4 minutes and 28 seconds for any one location. The eclipse will end at 20:52 UTC. However, the duration of total eclipse is slightly shorter, from 16:38 to 19:55 UTC, or 3 hours 17 minutes.

\section{Arago spot}
\begin{figure}[h]
    \centering \includegraphics[width=0.9\columnwidth]{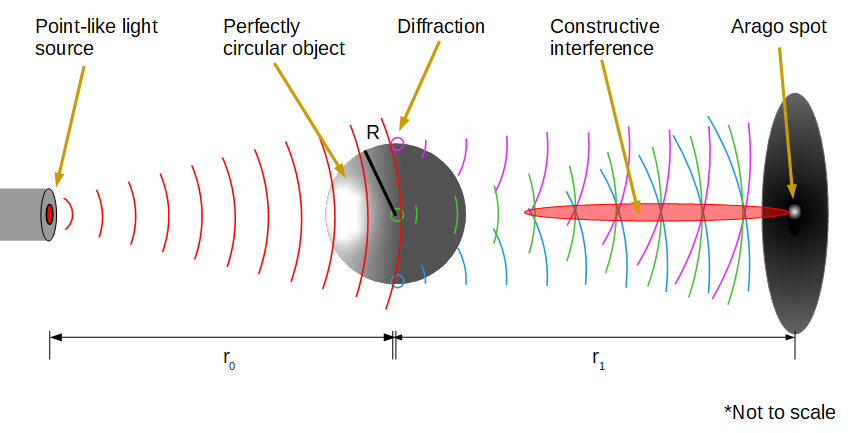}
    \caption{Diffraction around a perfect circle and the formation of the Arago spot}
    \label{fig:arago_formation}
\end{figure}

Arago spot~\cite{harvey}, also called Poisson spot or Fresnel spot, is an intensity maximum at the center of a shadow cast by a spherical object. It is created by the constructive interference of the diffracted waves originating from the area around the surface of the object. For a perfect sphere, the point right at the center of the shadow is at equal distance from all points at the rim of the cross section of the sphere, leading into an intensity maximum (Figure \ref{fig:arago_formation}). Arago spot is a result of the wave nature of the illuminating field, such as light, as was observed in 1818 by Arago and possibly a century earlier by Maraldi\cite{born&wolf}. While the study of diffraction patterns usually concerns visible light, De Broglie's wave nature of matter makes diffraction theory applicable for particles, such as neutrinos, as well.

\begin{figure}[h]
    \centering
    \includegraphics[width=0.6\columnwidth]{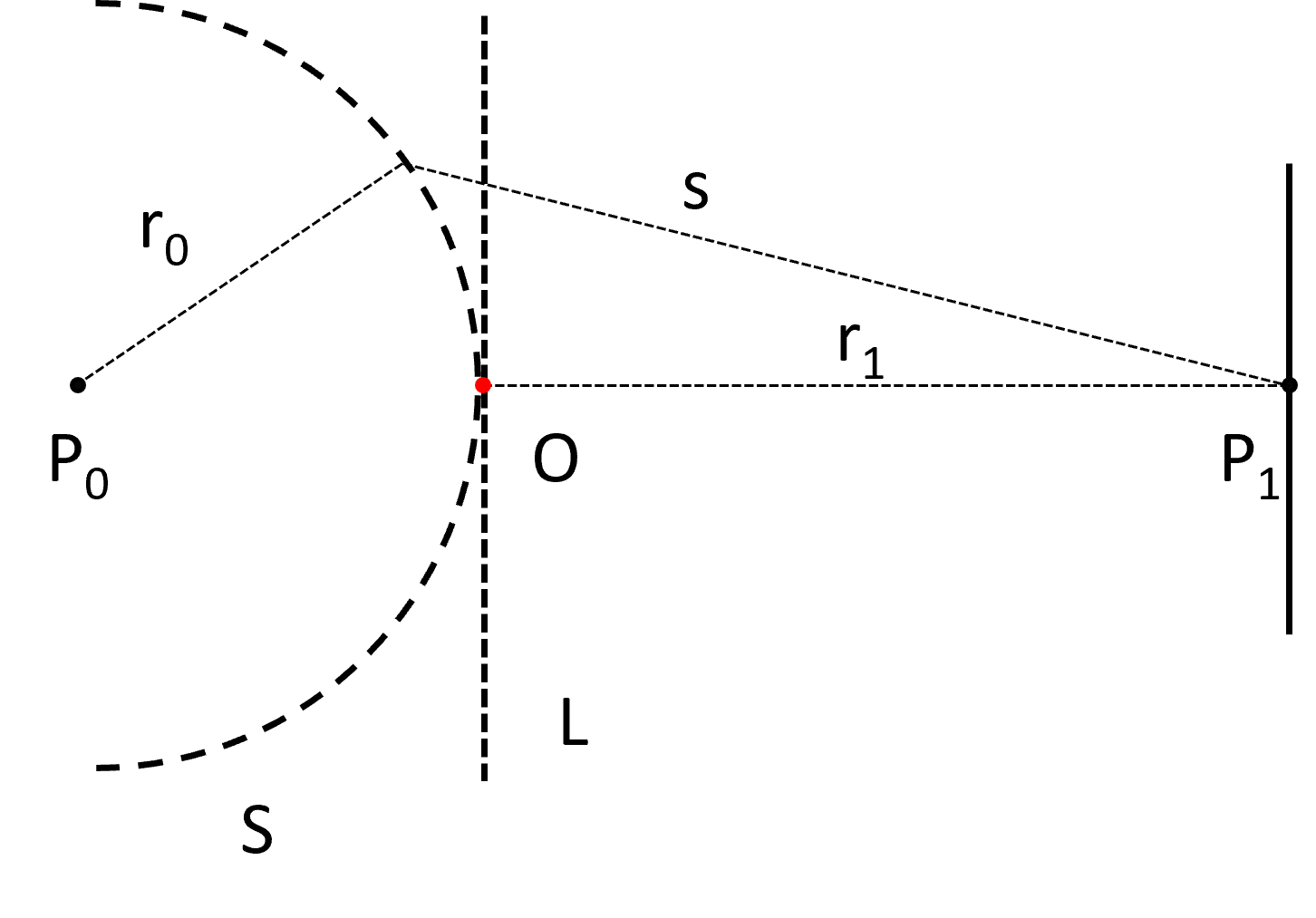}
    \caption{The geometry of the wave diffraction and the change of the surface integral from spherical surface $S$ to plane $L$.}
    \label{fig:integration-plane} 
\end{figure}

The Fresnel-Kirchoff diffraction formula for a point source gives as the amplitude of the diffracted field
\begin{equation}
U(P) = \frac{i}{\lambda} \frac{e^{ikr_0}}{r_0} \int_\mathcal{S} \frac{e^{iks}}{s}K(\chi) dS,
\end{equation}
where $\lambda$ is the wavelength of the light source, $r_0$ the distance between the source and the obstructing object, $k=\frac{2\pi}{\lambda}$ is the wave number, $dS$ is an infinitesimal surface element on the primary wavefront, $s$ the distance between the infinitesimal surface element $dS$ and the observed point on the screen and $K(\chi)$ the inclination factor. See Figure \ref{fig:integration-plane} for illustration. Following Kirchoff, assuming that both the distances from the source to the object and from the object to the screen are much larger than the size of the object itself, the inclination factor can be expressed as
\begin{equation}
K(\chi) = \frac{1}{2} \left[ 1+\cos \chi\right].
\end{equation}

The diffracted field $U(P)$ is dominated by contributions close to the surface of the object, as the secondary wavelets from more distant points are cancelled by both the decreasing inclination factor $K(\chi)$ and the rapidly oscillating phase factor $e^{iks}$.
Thus, we can approximate the integral over the spherical surface $S$ by a corresponding integral over a plane surface $L$ around the object.
The integration plane $L$ consists of unobstructed surface elements from distance $R$ around the object to infinity, as shown in Figure~\ref{fig:integration-plane}. With this transformation we have made the assumption that all secondary wavelets from the original source $P_0$ would be in phase at $L$. 

Following for example Reisinger et al.~\cite{poisson-intensity} we assume the inclination factor to be 1, yielding as the amplitude
\begin{equation}
U(P) = \frac{i}{\lambda} \frac{e^{ikr_0}}{r_0} \int_L \frac{e^{iks}}{s}dS,
\end{equation}
where the surface integral is now over the plane $L$.

Expressing the surface integral in polar coordinates, the distance $s$ from the surface element $dS$ at distance $\rho$ from the center of the object and in the azimuthal angle $\theta$ is
\begin{equation}
s = \sqrt{r_1^2 + \delta x^2 + \delta y^2} = \sqrt{r_1^2 + (r - \rho \sin \theta)^2 + (\rho \cos \theta)^2} = \sqrt{r_1^2 + r^2 - 2r\rho \sin \theta + \rho^2},
\end{equation}
where $r$ is the distance of the observation point $P$ from the center of the shadow. The amplitude becomes now
\begin{equation}
U(r) = \frac{i}{\lambda} \frac{e^{ikr_0}}{r_0} \int_{R}^\infty d\rho \int_0^{2\pi} d\theta\, \rho \frac{e^{ik\sqrt{r_1^2 + r^2 - 2r\rho \sin \theta + \rho^2}}}{\sqrt{r_1^2 + r^2 - 2r\rho \sin \theta + \rho^2}}.
\end{equation}

Linearizing the square root yields
\begin{equation}
\sqrt{r_1^2 + r^2 - 2r\rho \sin \theta + \rho^2} \approx r_1  + \frac{r^2 - 2r\rho \sin \theta + \rho^2}{2r_1}
\end{equation}
and the integral becomes
\begin{equation}
U(r) = \frac{i}{\lambda} \frac{e^{ik(r_0+r_1)}}{r_1 r_0} e^{ik\frac{r^2}{2r_1}}\int_{R}^\infty d\rho \int_0^{2\pi} d\theta\, \rho e^{ik\frac{\rho^2- 2r\rho \sin \theta}{2r_1}}
\end{equation}
or, in terms of wavelength $\lambda = 2\pi/k$,
\begin{equation}
U(r) = \frac{i}{\lambda} \frac{e^{ik(r_0+r_1)}}{r_1 r_0} e^{ik\frac{r^2}{2r_1}}\int_{R}^\infty d\rho\, \rho e^{i\pi\frac{\rho^2}{\lambda r_1}} \int_0^{2\pi} d\theta\, e^{-i\pi\frac{2r\rho \sin \theta}{\lambda r_1}}.
\end{equation}
The angular integral is the zeroth order Bessel function $2\pi J_0(y)$, where the argument $y = \pi\frac{2r\rho}{\lambda r_1}$.
We get for the amplitude
\begin{equation}
U(r) = \frac{2\pi i}{\lambda} \frac{e^{ik(r_0+r_1)}}{r_1 r_0} e^{i\pi \frac{r^2}{\lambda r_1}}\int_{R}^\infty d\rho\, \rho e^{i\pi\frac{\rho^2}{\lambda r_1}} J_0(\pi\frac{2r\rho}{\lambda r_1}).
\end{equation}
By doing change of variable $x = \rho^2$ and writing the constant expression in front of the integral as $u_0$ for brevity we get

\begin{equation}
U(r) = u_0 \frac{1}{2}\int_{R^2}^\infty dx\, e^{i\pi\frac{x}{\lambda r_1}} J_0(\pi\frac{2r\sqrt{x}}{\lambda r_1}).
\label{eq:diffracted_field}
\end{equation}

Next we introduce a convergence factor $e^{-\eta x}$ into the integral to ensure it converges. The physical rationale for including the convergence factor is that surface elements with a vertical component that is largely different from the observation point $P$ contribute little to the aggregated amplitude at $P$ and therefore their impact can be discarded. We get
\begin{equation}
U(r) = u_0 \frac{1}{2}\int_{R^2}^\infty dx\, e^{i\pi\frac{x}{\lambda r_1}} e^{-\eta x} J_0(\pi\frac{2r\sqrt{x}}{\lambda r_1})
\end{equation}
where $\eta$ is a convergence parameter. In the limit of small value of $\eta$, we regain the original diffracted field in Eq.~\eqref{eq:diffracted_field}.

\begin{figure}[ht]
    \centering
    \includegraphics[width=0.9\columnwidth]{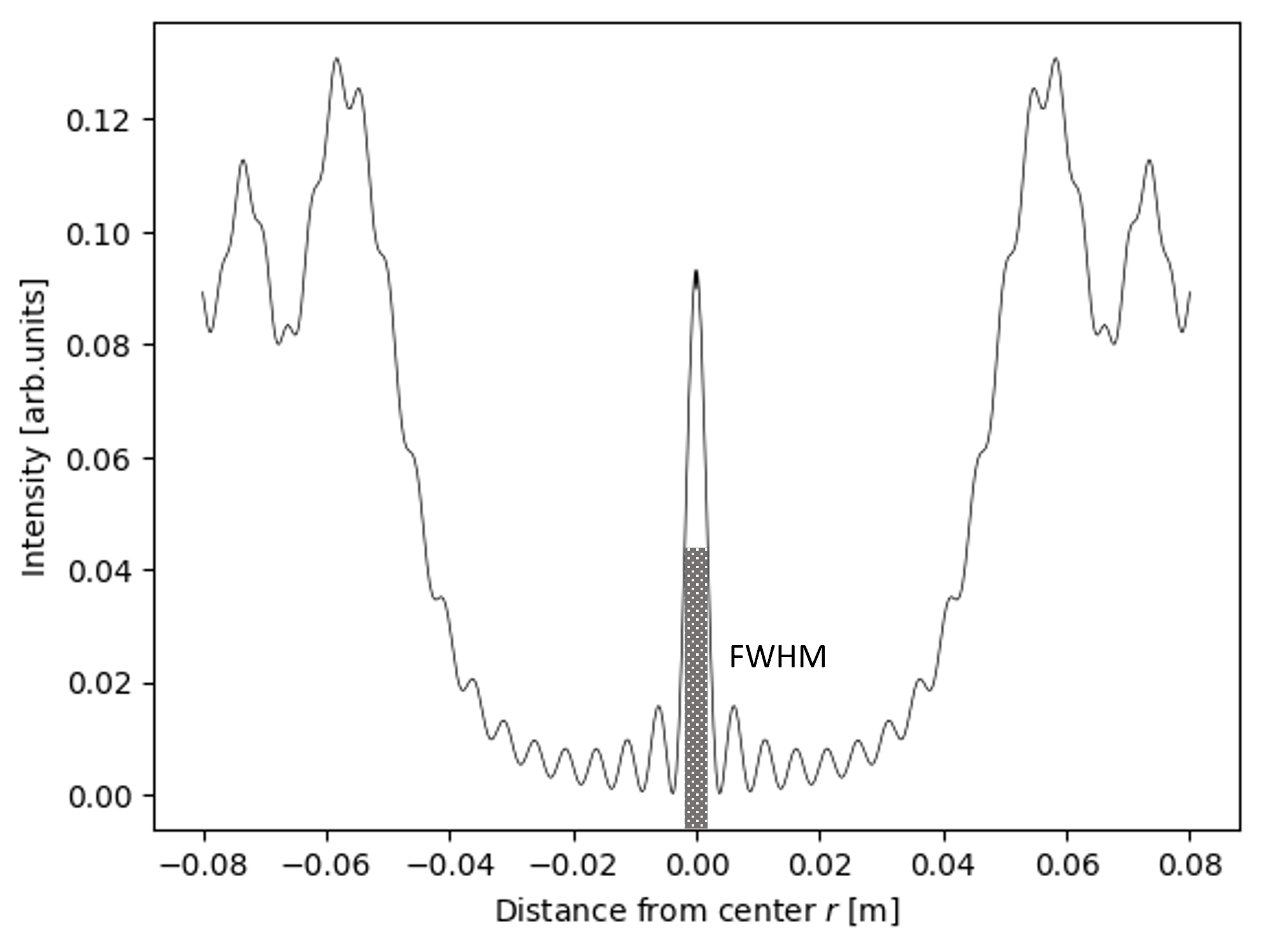}
    \caption{The intensity distribution $|U(r)|^2$ of the diffraction pattern on the screen as a function of the distance from the center. To achieve the illustrated pattern, we used the parameters $\eta$ = 0.002, $\lambda$ = $1\cdot 10^{-12}$ m, $r_1 = 400,000\,\mathrm{km}$, $R = 0.04\,\mathrm{m}$. Shown is also the full width at half maximum (FWHM) of the central intensity maximum. Notice that these parameters do not correspond to the neutrino diffraction around the Moon as the obstructing object radius is here smaller by a factor $4\cdot 10^7$.}
    \label{fig:bessel-figure} 
\end{figure}

Evaluating the above integral numerically for a range of values for $r$ (the distance from the center of the screen) we get a diffraction pattern with the shape illustrated in Figure~\ref{fig:bessel-figure}. The figure shows a typical diffraction pattern with the Arago spot in the middle, followed by regions alternating between shadow and medium intensity as we traverse further from the center. The distribution shows that the intensity maximum at the Arago spot is close to the intensity in the regions that are not directly obstructed by the object between the screen and the source. This result is expected (see for example Born and Wolf (1999)\cite{born&wolf}) and proves that our simplifying assumptions for the diffraction profile are reasonable. 

The full width at half maximum (FWHM) for the theoretically observed pattern is between r = -0.00184 m and r = 0.00184 m for a total width of 0.00368 m, i.e. approximately 3.7 mm for the width of the Arago spot. 
However, the parameters used in the depiction of the intensity distribution in Figure~\ref{fig:bessel-figure} do not completely correspond to the case of the solar neutrino flux around the Moon since numerical analysis for an obstructing object with radius of $1700\,\mathrm{km}$ is impractical. However, the Arago spot is a robust feature and analytically its width can be shown to be~\cite{emile-emile}
\begin{equation}
\delta \sim \frac{\lambda r_1}{R}
\label{eq:arago_width}
\end{equation}
where $\lambda$ is the wavelength, $r_1$ is the distance between the obstructing object and the screen and $R$ the radius of the obstructing object.

Arago spot is very sensitive to small imperfections in the smoothness of the cross-section obstructing object and  the required surface smoothness follows the same scaling as the width of the spot, being
\begin{equation}
\delta R \leq \frac{\lambda r_1}{R},
\end{equation}
where $\delta R$ is the deviation from the perfect sphere with constant radius $R$.

The smoothness requirement explains why the Arago spot is rarely seen in the real world outside the laboratory. For example, for visible light arriving from the Sun and obstructed by the Moon ($\lambda = 500\mathrm{nm}$, $r_1 = 400,000\,\mathrm{km}$ and $R = 1,740\,\mathrm{km}$), the maximum allowed smoothness deviation would be approximately $\delta R \leq 100\,\mathrm{\mu m}$ which obviously does not agree with reality.

\section{Solar neutrino diffraction}

Solar neutrinos are neutrinos which are created in the fusion reactions of the Sun~\cite{solarneutrinos}. Almost all solar neutrinos are produced in the proton-proton chain or p-p chain for short. The reactions and energies of the corresponding neutrinos are as follows:
\begin{align*}
p+p&\longrightarrow \leftindex^2 {H}+e^{+}+v_e\; (E_v < 0.42 \text{MeV}) \\
p+e^-+p&\longrightarrow \leftindex^2 {H}+v_e\; (E_v = 1.44 \text{MeV}) \\
\leftindex^7 {Be}+e^-&\longrightarrow \leftindex^7 {Li}+v_e\; (E_v = 0.86\text{MeV}\: (90\%),\: 0.38\text{MeV}\: (10\%)) \\
\leftindex^8 {B}&\longrightarrow \leftindex^8 {Be}+e^+ +v_e\; (E_v < 15\text{MeV}) \\
\leftindex^3 {He}+p&\longrightarrow \leftindex^4 {He}+e^+ +v_e\; (E_v < 18.8\text{MeV}) \\
\end{align*}
However, the two most energetic neutrinos in the chain contribute to only a really small portion of the total neutrino flux. For the sake of this research, it is fair to assume that the average energy of a solar neutrino is $E\approx\text{0.5 MeV}$. \\

The de Broglie relation connects the energy $E$ and de Broglie wavelength $\lambda_\mathrm{dB}$of the solar neutrino 
\begin{equation}
 E = \frac{hc}{\lambda_\mathrm{dB}},
\end{equation}
and hence the typical solar neutrino wavelength is
\begin{equation}
\lambda_\mathrm{dB} = \frac{hc}{E} \approx \frac{7\cdot 10^{-34}\,\mathrm{Js} \cdot 3 \cdot 10^8 \,\mathrm{m/s}}{0.5\,\mathrm{MeV}} \approx 2.5\,\mathrm{pm}.
\end{equation}
Detecting the solar neutrino Arago spot would determine the position of the center of the Moon's shadow, and hence the position of the Moon, with the precision given by the width of the Arago spot in Eq.~\eqref{eq:arago_width}
\begin{equation}
\delta = \frac{r_1 \lambda_\mathrm{dB}}{R} = \frac{4 \cdot 10^8\,\mathrm{m} \cdot 2.5 \cdot 10^{-12}\,\mathrm{m}}{1.7 \cdot 10^6\,\mathrm{m}} \approx 0.6\,\mathrm{nm},
\end{equation}
where the Earth-Moon distance $r_1 = 400 000\,\mathrm{km} = 4\cdot 10^{8}\,\mathrm{m}$, and the Moon radius $R=1700\,\mathrm{km}$.

The mean free path of a solar neutrino in lead can be roughly estimated\cite{free-path} to be around 1000 light years.
If we approximate the same mean free path for solar neutrinos traveling through moon, the fraction of solar neutrinos interacting with the Moon when traveling through the center would be around
\begin{equation}
\alpha \approx \frac{2 \cdot 10^6\text{m}}{1000\,\text{ly}} \approx 2 \cdot 10^{-13},
\end{equation}
which means that the Moon is basically transparent for neutrinos. If we multiply this fraction with the total rate of neutrinos passing through the Moon
\begin{equation}
f_\text{pass} = \sigma \pi R^2
\end{equation}
where $R = 10^6 \mathrm{m}$ is the radius of the Moon and $\sigma = 7 \cdot 10^{14}\,\mathrm{1/(m^2s)}$ is the total neutrino flux from Sun observed at Earth, we get
\begin{equation}
f_\text{diffract} = \alpha f_\text{pass} = \alpha \sigma \pi R^2 \approx 10^{15} \, \mathrm{1/s}
\label{eq:diffracted_fraction}
\end{equation}
which is the total rate of diffracting neutrinos.

\begin{figure}[H]
    \centering
    \includegraphics[width=0.8\columnwidth]{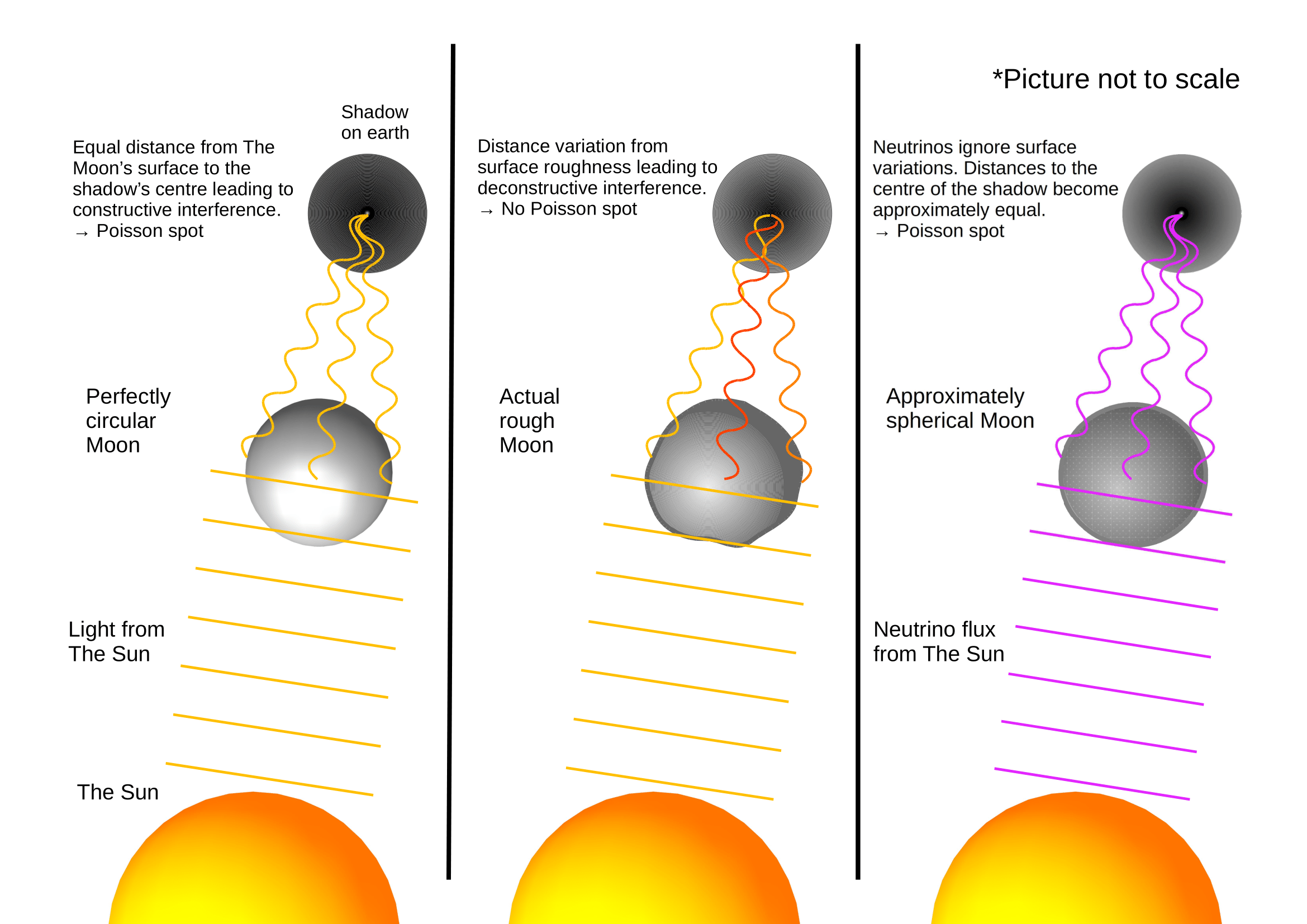}
    \caption{Solar neutrino diffraction from the Moon}
    \label{fig:neutrino_vs_light}
\end{figure}

Since there is a small portion of the neutrinos that do interact with moon, they could form an Arago spot during a solar eclipse. The roughness of the Moon that made Arago spot impossible for light is not a problem for solar neutrinos due to their ability to pass through matter (Figure \ref{fig:neutrino_vs_light}). For example, a crater on the surface of the Moon causing a 100m local variance in the radius would interact approximately with only
\begin{equation}
\frac{100\text{m}}{1000\text{ly}} \approx 10^{-17}
\end{equation}
of the total neutrinos passing through the Moon, which is less than $10^{-4}$ of the total diffracting neutrinos.

Since the fraction of solar neutrinos interacting with the Moon is so small, it is reasonable to consider, whether the increase in solar neutrinos detected at the Arago spot is enough to be differentiated from the surroundings. 
The intensity maximum at the Arago spot approaches the corresponding unobstructed intensity in the far field. The contrast between the central intensity and the full shadow, compared with the background neutrino flux that passes through the Moon undiffracted is only $10^{-13}$, meaning that in practice the solar neutrino Arago spot is unobservable.

\section{Entanglement and detecting the position of the Moon}

Entangled particles are correlated in a way that makes it possible to deduct properties of one by making observations of the other as postulated by Einstein-Podolsky-Rosen\cite{epr} (EPR). The diffracted neutrinos become entangled with the mass center of the Moon, evoking a connection between the two that can be characterized by Heisenberg's uncertainty principle.

Heisenberg's uncertainty relation states that Nature has a minimum threshold for uncertainty given by
\begin{equation}
\delta p \delta x \geq \hbar
\end{equation}
where $\delta p$ is the uncertainty in momentum, $\delta x$ the uncertainty in position and $\hbar$ Planck's constant.
Now if we could observe the Arago spot created by the neutrinos diffracted from the Moon during a solar eclipse, we could infer the location of the Moon with $1$ nm precision. Given this observation made by the mere existence of the Arago spot, the uncertainty principle states that the uncertainty in the Moon's momentum would have to be
\begin{equation}
\delta p \geq \frac{\hbar}{\delta x} = \frac{10^{-34}\,\mathrm{Js}}{10^{-10}\,\mathrm{m}} = 10^{-24}\,\mathrm{kgm/s}.
\end{equation}
Given the Moon's mass of $7\cdot 10^{22}\,\mathrm{kg}$, the change in velocity would be $10^{-46}\,\mathrm{m/s}$ for a single neutrino observation.

Multiple neutrino observations will lead into a random walk in momentum, with the total change in the momentum scaling as square root of the number of observations. That is,
\begin{equation}
\delta P = \sqrt{N} \delta p.
\end{equation}
The total diffracted neutrino rate was calculated in Eq.~\eqref{eq:diffracted_fraction} as $f=10^{15}\,\mathrm{1/s}$ and the solar eclipse in April lasts for roughly three hours (approximated as 10,000 seconds).
Thus the total number of neutrinos (observations, $N$) contributing to the neutrino Arago spot during the whole eclipse is on the magnitude of $10^{19}$. Observation (in this case, 1000 light years of lead will be sufficient to conduct the measurement) of the solar neutrino Arago spot during the whole eclipse will lead into total momentum change of
\begin{equation}
\delta P = \sqrt{10^{19}} \cdot 10^{-24}\,\mathrm{kgm/s} \approx  10^{-15}\,\mathrm{kgm/s}.
\end{equation}

However, only a fraction of the diffracted neutrinos will be diffracted to the Arago spot, making the above estimate a worst case scenario for the Moon. To stress test the Moon even further, we can consider the outlandish assumption that all the neutrinos impinging on the Moon contributed to the spot. That would imply a rate of
\begin{equation}
\pi R_{moon}^2 \sigma \approx 2 \cdot 10^{27}\,\mathrm{1/s}
\end{equation}
or $2 \cdot 10^{31}$ neutrinos in total during the whole eclipse. Still, that would give only a total momentum change of
\begin{equation}
10^{-24}\,\mathrm{kgm/s} \cdot \sqrt{2 \cdot 10^{31}} \approx 4 \cdot 10^{-9}\,\mathrm{kgm/s}.
\end{equation}
The above is based on the assumption of random walk in the momentum changes. With perfect uniform orientation we would get momentum change of
\begin{equation}
10^{-24}\,\mathrm{kgm/s} \cdot 2 \cdot 10^{31} = 2 \cdot 10^{7}\,\mathrm{kgm/s}
\end{equation}
which would correspond to a change in velocity of only $10^{-15}\,\mathrm{m/s}$ which would not be sufficient to set the Moon free. For a sense of scale, a 1000 kg asteroid moving at $20\,\mathrm{km/s}$ carries a similar amount of momentum. 

Furthermore, the probability of such event can be approximated as follows. By assuming that all changes in momentum happen in a direction less than a $12\degree=\pi/15$ difference from a certain trajectory we can calculate the probability for a single change in momentum to happen in that direction. The probability of such event is simply equal to the area of a sector of a sphere divided by the total area of the sphere.
\begin{equation}
p=\frac{A_\mathrm{sector}}{A_\mathrm{sphere}}=\frac{1}{4\pi R^2}\int_0^{\frac{\pi}{15}}\int_0^{2\pi}2\pi R^2\mathrm{sin\varphi}d\varphi d\theta  \approx 1\ \%.
\end{equation}
If we assume that all $2\cdot 10^{31}$ neutrinos contribute to the change of direction in the 12\degree\:range mentioned earlier, the probability for that would be 
\begin{equation}
p=10^{-2\cdot10^{31}}
\end{equation}
which is ridiculously small.

\section{Conclusions}

Our analysis shows that measuring the Moon during a solar eclipse poses no risk for the heavenly body. We calculate that even in the unlikely worst-case scenario, observing the solar Arago spot would only result in the Moon's momentum shift similar to that caused by an asteroid the mass of a saltwater crocodile hitting the lunar surface. Therefore the idea of unobserving the Moon will continue as an area which requires no further study.

\end{document}